\documentclass[useAMS,usenatbib,fleqn]{mn2e}
\usepackage{subfigure}
\usepackage{graphicx, amssymb}
\usepackage[fleqn]{amsmath}
\usepackage{epsfig}
\usepackage{color}
\usepackage{times}

\title[The First Low-Mass Stars]{Formation of the First Low-Mass Stars from Cosmological Initial Conditions}

\author[C. Safranek-Shrader et al.]
{Chalence~Safranek-Shrader, Milo\v s~Milosavljevi\'c, and Volker~Bromm\\
Department of Astronomy, University of Texas at Austin, Austin, TX 78712, USA
 }
 
\newcommand{\nh}{n_{\mathrm{H}}}
\newcommand{\kelvin}{\mathrm{K}}
\newcommand{\cc}{\mathrm{cm}^{-3}}
\newcommand{\msun}{M_{\odot}}

\newcommand{\zsun}{Z_{\odot}}

\newcommand{\htwo}{\mathrm{H}_2}

\newcommand{\tcmb}{T_{\mathrm{CMB}}}

\newcommand{\kb}{k_{\mathrm{B}}}

\newcommand{\pc}{\mathrm{pc}}
\newcommand{\au}{\mathrm{AU}}

\newcommand{\yr}{\mathrm{yr}}

\newcommand{\cs}{c_{\mathrm{s}}}

\newcommand{\lj}{L_{\mathrm{J}}}

\newcommand{\mj}{M_{\mathrm{J}}}

\newcommand{\racc}{r_{\mathrm{acc}}}
\newcommand{\msunperyr}{M_{\odot}\,\mathrm{yr}^{-1}}

\newcommand{\tdust}{T_{\mathrm{d}}}
\newcommand{\betaesc}{\beta_{\mathrm{esc}}}

\newcommand{\taucont}{\tau_{\mathrm{cont}}}

\begin{document}

\label{firstpage}

\maketitle
\topmargin-1cm

\begin{abstract}
We simulate the formation of a metal-poor 
($10^{-2}\,\zsun$) stellar cluster in one of the first galaxies to
form in the early Universe, specifically a high-redshift 
atomic cooling halo ($z\sim14$). 
This is the first calculation that resolves the
formation of individual metal-enriched stars 
in simulations starting from realistic cosmological initial conditions.  We
follow the evolution of a single dense clump among several in the
parent halo. The clump 
forms a cluster of $\sim40$ stars and sub-stellar objects within $7000$ yrs 
and could continue forming stars $\sim5$ times longer.
Protostellar dust heating has a negligible effect on the
star formation efficiency, at least during the early evolutionary
stages, but it moderately suppresses gaseous
fragmentation and brown dwarf formation. 
We observe fragmentation in
thin gaseous filaments and sustained accretion in larger, rotating
structures as well as ejections by binary interactions.
 The stellar
initial mass function above
$0.1\,\msun$, evaluated after $\sim10^4\,\yr$ of fragmentation and
accretion, seems in agreement with the recent measurement 
in ultra-faint dwarf spheroidal Galactic satellites of \citet{Geha13}. 
\end{abstract}


\section{Introduction}
\label{sec:intro}
Observations of low-metallicity stars in the stellar halo and
satellites of the Milky Way have furnished an expanding window
into how the Galaxy's primitive precursors formed and evolved in the
ancient Universe.  In a pursuit named ``stellar archaeology,'' the
 chemical abundance patterns of metal-poor
stars are used to expose the character of the supernovae that had
enriched them and the physical state of their formation environment
\citep[e.g.,][]{Frebel05,Karlsson13}. In particular, 
the metallicities and abundance patterns in the faintest and most
metal-poor galaxies, the ultra-faint dwarf spheroidal satellites
(UFDs), can potentially be used to probe the very first stellar
generation forming from metal-free initial conditions and its immediate
successors \citep{Brown12,Frebel12,Vargas13}. Furthermore, there are hints 
that in the measured mass range, the stellar initial mass function (IMF) is shallower
 in UFDs than in more metal-rich and evolved galaxies \citep{Geha13}, a potential 
 challenge to star formation models and an important clue as to the origin of UFDs and 
 of primitive, early stellar systems in general.\\
\indent Capitalizing on the information provided by the fossils of the first galaxies requires 
a theoretical understanding
of star formation under conditions distinct from those in the present 
Milky Way \citep[e.g.,][]{Bromm13}. Typical metallicities in UFDs are 
only $\sim1\%$ of the solar value \citep[e.g.,][]{Kirby11}. Current theory 
suggests that the first metal-enriched stellar generations formed in predominantly 
atomic gas clouds assembled by dark-matter-driven
cosmic infall and thermal instability, and were subject to a
temperature floor imposed by the cosmic microwave background (CMB)
\citep[e.g.,][]{Wise12,SafranekShrader14}. In contrast, the bulk of
present-day star formation, where recent numerical breakthroughs have
focused \citep[e.g.,][]{Bate12,Krumholz12,Federrath12}, occurs in turbulent
molecular clouds where the thermodynamics is dominated by dust and its
coupling to the interstellar radiation field. Salient aspects of star
formation in molecular clouds could apply to
all systems with supersonic gas flow velocities, excluding
star formation at extremely low or zero metallicities that operates 
in a distinct, Population~III mode \citep[e.g.,][]{Yoshida06,Clark11,Greif11,Dopcke13}.
These aspects include the formation of low-mass stars through
turbulent gravitational fragmentation and of more
massive stars through the coherent collapse of self-gravitating cores
as well as (possibly ``competitive'') accretion stabilized by local protostellar radiative heating \citep{Bate03,MacLow04,McKee07,Zinnecker07}.\\
\indent Our goal is to investigate the formation of the
first generation of metal-enriched stars from cosmological initial
conditions so that we can begin charting out the UFD formation history, relating it 
to concepts normally used to describe star formation under distinct, molecular-cloud-like conditions.
We present highly zoomed adaptive-mesh-refinement (AMR) hydrodynamical
simulations extending those in \citet{SafranekShrader14} to much
higher resolution, resolving the density and length scales where
protostellar masses are imprinted. Namely, gaseous collapse in the
presence of dust is now tracked to the densities
$\sim10^{13}\,\textrm{cm}^{-3}$ where gas becomes optically thick to
its cooling radiation and further fragmentation is strongly suppressed.
We use Lagrangian sink particles to follow protostellar accretion after the initial collapse.\\
\indent Importantly, the simulations here are derived from a section of a 
coarser simulation initialized from a realization of the $\Lambda \mathrm{CDM}$ 
cosmological model, instead of beginning from an artificially generated turbulent 
velocity field and spherically symmetric gas configuration. Therefore the gas flow 
morphology in the simulations here is a self-consistent outcome of turbulent 
virialization in a dark matter halo, thermal instability, and gravitational collapse.

\section{Methodology}
\label{sec:method}

The simulations were performed with the AMR hydrodynamics code
\textsc{flash} \citep{Fryxell00}, version 4. The initial conditions
were extracted from the highest metallicity cosmological simulation of
\citet{SafranekShrader14}.  The original cosmological simulation was
performed in a $1\,\textrm{Mpc}$ comoving box with standard
$\Lambda$CDM metal-free initial conditions and an externally imposed
$\htwo$-dissociating UV background that prevented star formation in
halos that have not reached the atomic cooling limit. Upon identifying an atomic
cooling halo at $z=13.8$, we endowed gas within its virial radius
with a nonzero, uniform metallicity of $10^{-2}\,\zsun$, crudely
mimicking enrichment by preceding Population~III stars. The cooling
by metallic fine structure lines induced localized gaseous collapse in
the halo. Sink particles (hereafter ``sinks'') were allowed to form at
densities $>10^6\,\cc$ and had accretion radii $\sim10^4\,\au$. Since
fragmentation at still higher densities is
%and shorter length scales is
expected on physical grounds, the sinks in
\citet{SafranekShrader14} did not represent individual stars, but
pre-stellar clumps poised to form small stellar associations or
clusters \citep{Bergin07}.

To study the formation and evolution of
\emph{individual} stars, much higher resolution is required to attain
conditions at which the gas becomes optically thick to all forms of cooling
radiation and further fragmentation is thermodynamically suppressed (unresolved
fragmentation may still be possible in rotationally-supported
protostellar disks). At the metallicity considered here and in the
presence of dust, this occurs at a density $\sim10^{12}\,\cc$. Since
achieving the requisite dynamical range in a full cosmological
simulation is computationally prohibitive, we opted for the
``cut-out'' strategy, similar to \citet{Greif11}. When the maximum
density in the $10^{-2}\,\zsun$ run of \citet{SafranekShrader14}
reached $10^7\,\cc$, we extracted a cubical section of size
$0.52\,\pc$ containing a total gas mass of $390\,\msun$ centered on
the densest cell. The cut-out region contains a single, slightly
ellipsoidal pre-stellar clump undergoing supersonic compression along one
direction. The compression was produced by a collaboration of
gravitational and pressure forces in the
aftermath of thermal instability at densities
$\sim10^2-10^4\,\cc$ in the parent simulation.

Since the gravitational potential was strongly gas-dominated we
neglected the dark matter. We proceeded to integrate the cut-out
simulation with a Jeans length resolution of at least $24$ grid cells and 
subject to reflective hydrodynamical boundary
that represented an approximately pressure-confined environment.  
However, we were careful to run the
simulation only for a time much shorter than the sound
crossing time from the box center to the boundary ($\approx 0.4\,\textrm{Myr}$ 
at temperature $T=50\,\kelvin$), rendering the
nature of the boundary condition immaterial.  The gravitational
potential of the gas was obtained with the multigrid solver with
isolated gravitational boundary conditions. The simulations did not
include magnetic fields, though we plan to explore their significance
in future work.
%but if the gas metallicity is spatially
%uniform as assumed here, the magnetic fields injected by the metal
%donor Population~III stars would have been further amplified by the
%same process that homogenized the metals, potentially rendering them 
%dynamically important.

  \begin{figure}
%\includegraphics[scale=0.5, clip, trim = 20 5 5 5 ]{fig2}
%\begin{center}
\includegraphics[width=0.49\textwidth]{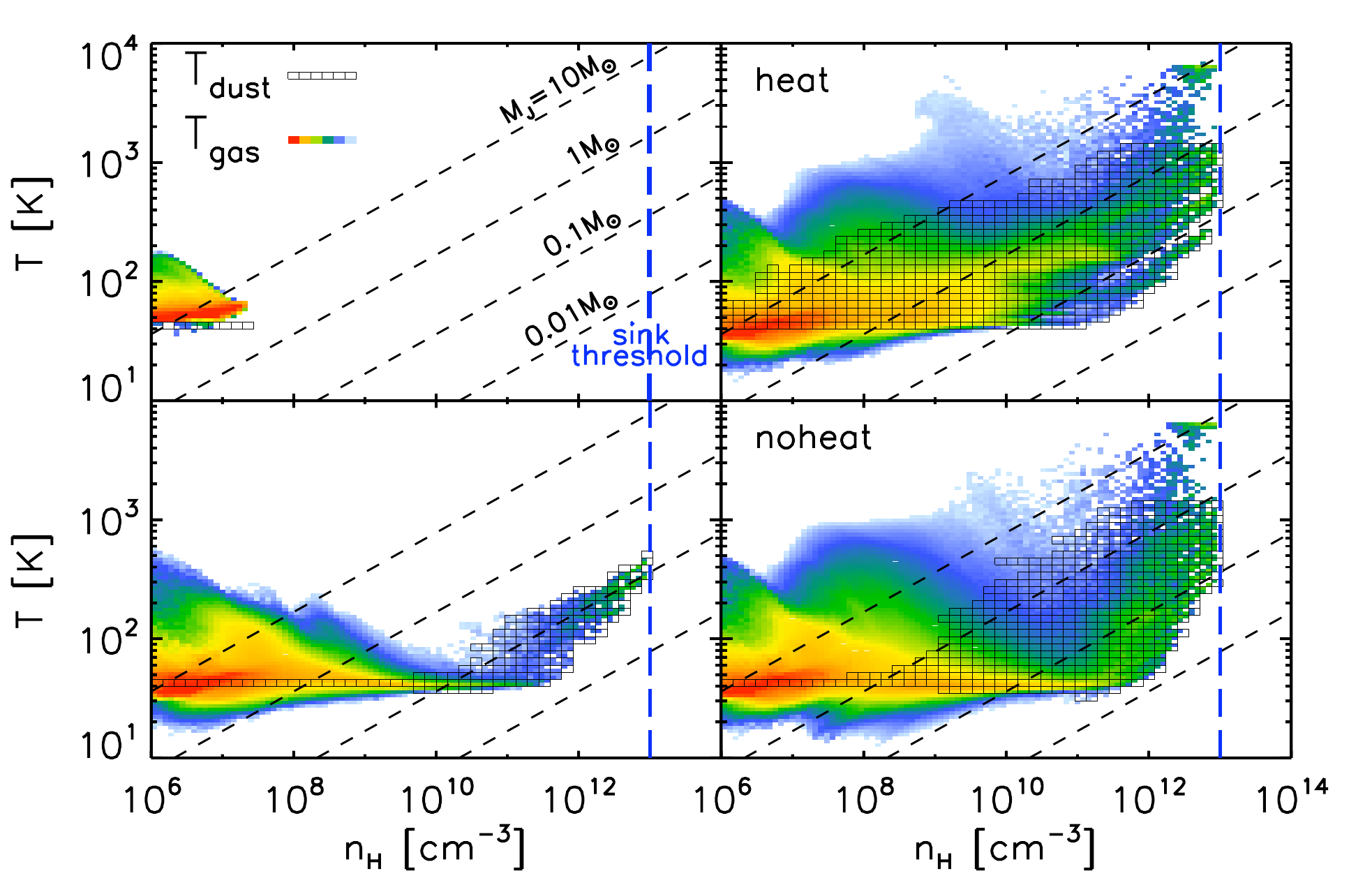}
\caption{Density-temperature distribution of the gas (color from white to red 
scaling with gas mass) and 
dust (boxed points).  Counterclockwise from top left, the panels show
  the beginning of the simulation, the onset of sink particle
  formation, and $7000\,\textrm{yr}$ later in the simulations
  \textsc{noheat} and \textsc{heat}. Dashed diagonals show
  lines of constant Jeans mass.}
\label{fig:dens_temp}
%\end{center}
\end{figure}

\renewcommand{\arraystretch}{0.4}
\begin{figure}
%\begin{center}
 \begin{tabular}{l l }
 % \imagetop{ \includegraphics[scale=\ascale, clip, trim = 0 0 0 0 ]{cut_nofeed_big_dens_xy_0193} }&
 \includegraphics[width=1.65in]{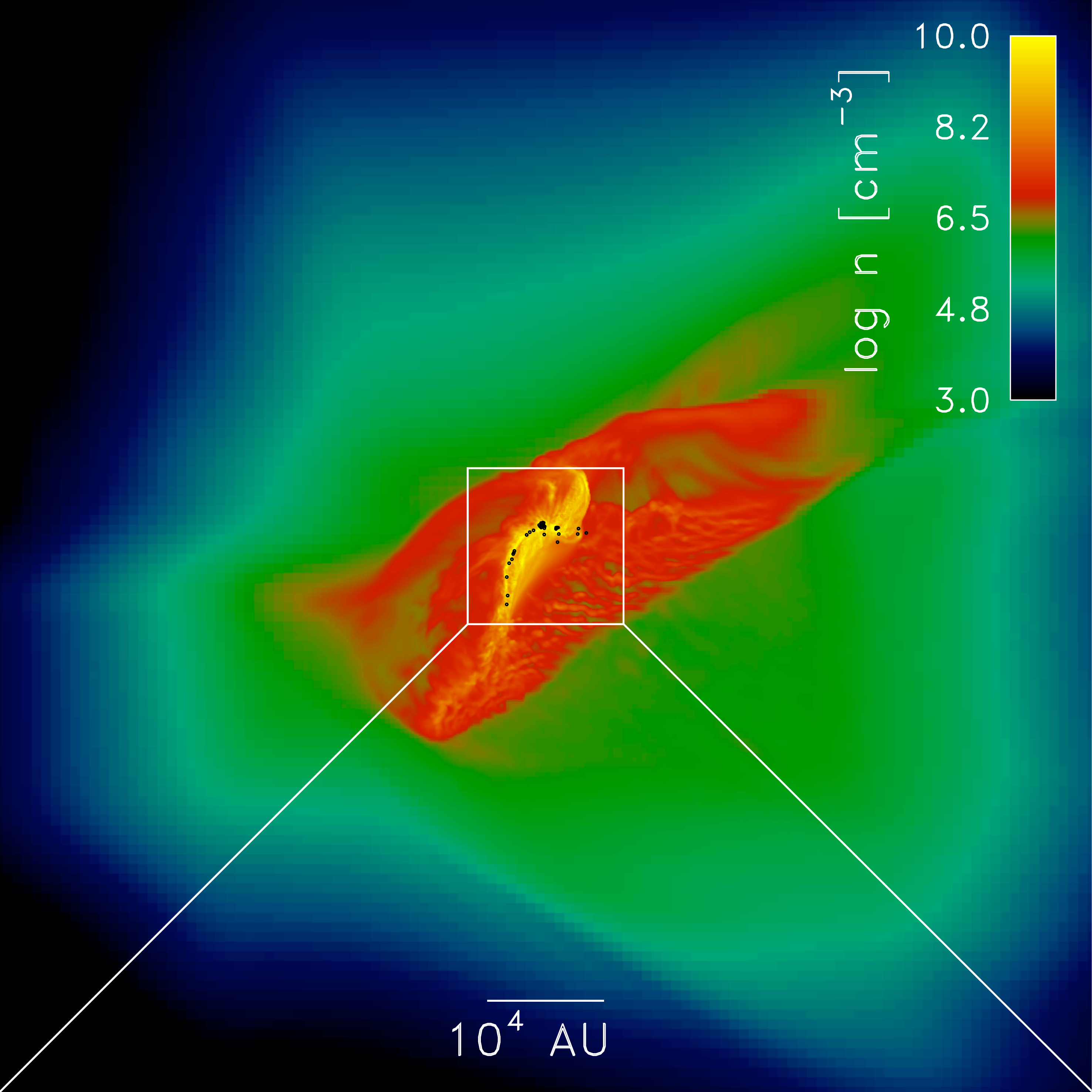} \hspace{0.01em}  &
  \includegraphics[width=1.65in]{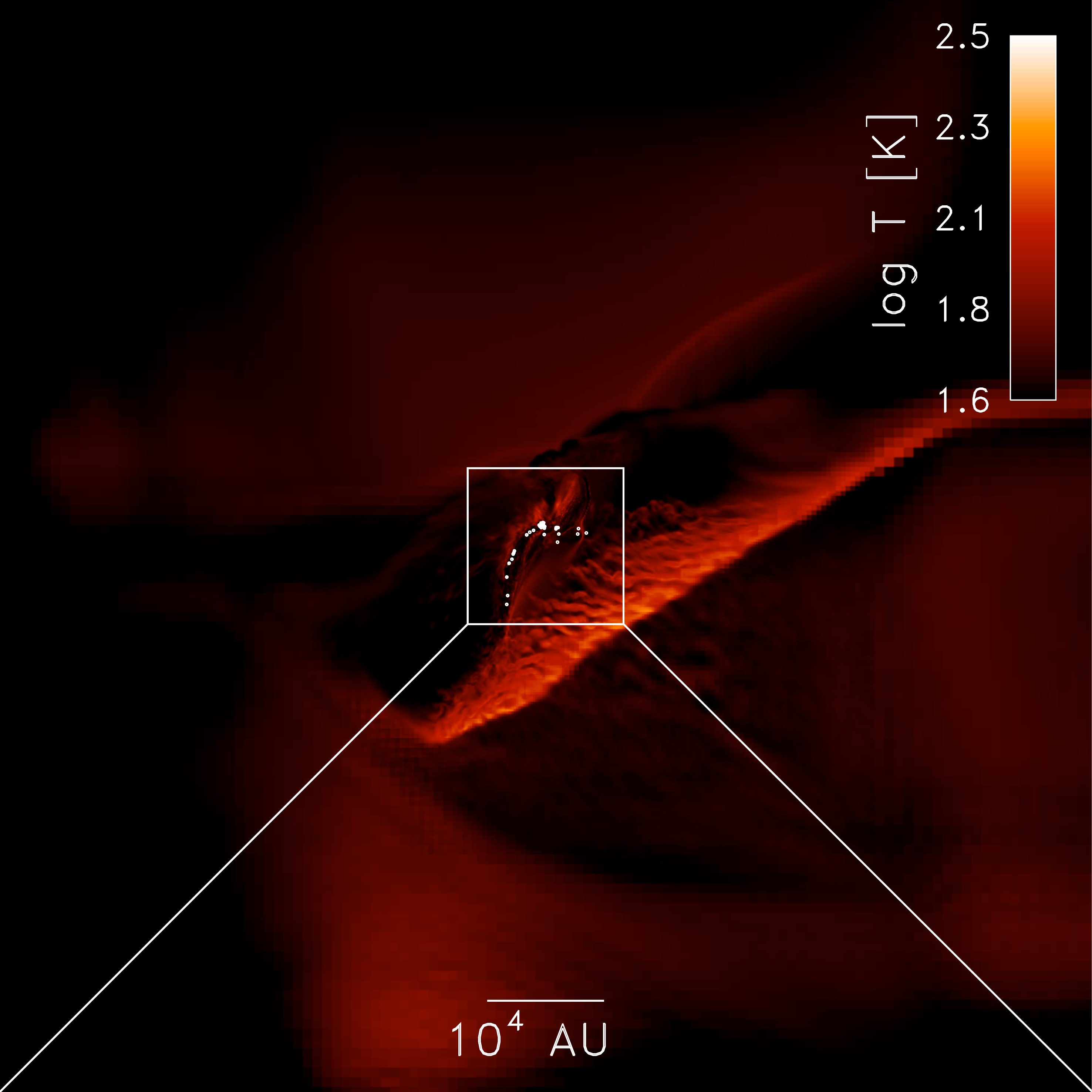}  \\
 \includegraphics[ width=1.65in]{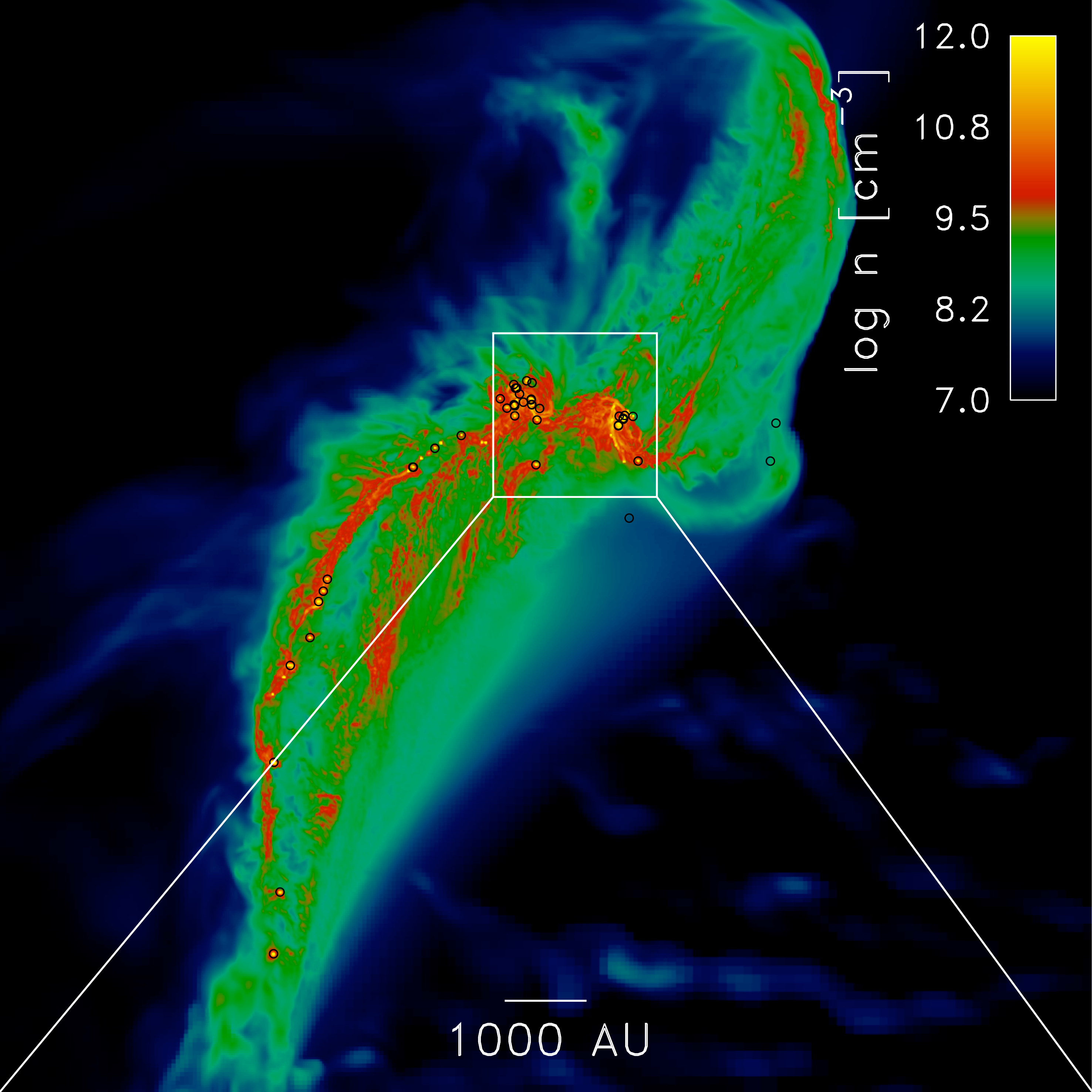} \hspace{0.01em}   &
 \includegraphics[width=1.65in]{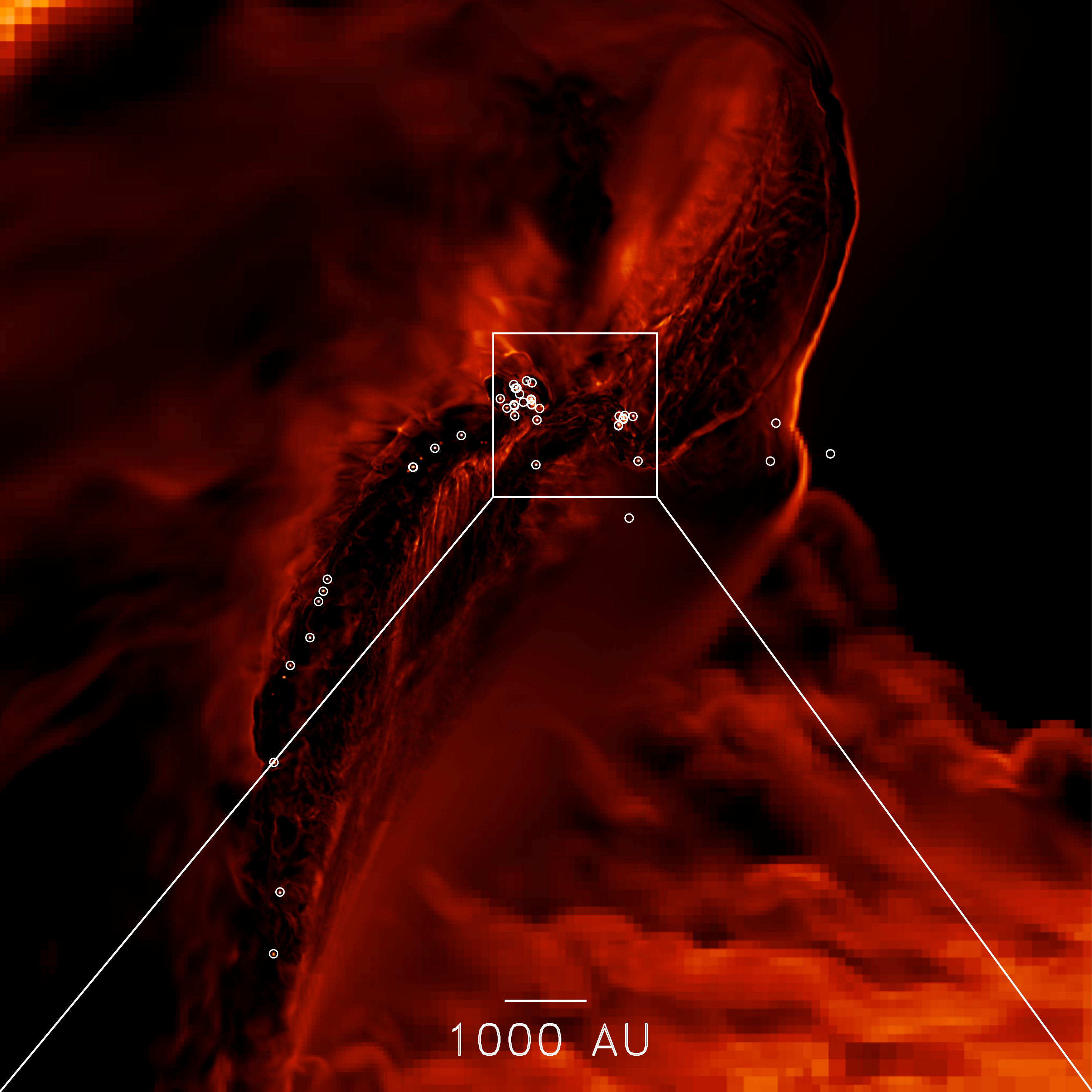}  \\
  \includegraphics[ width=1.65in]{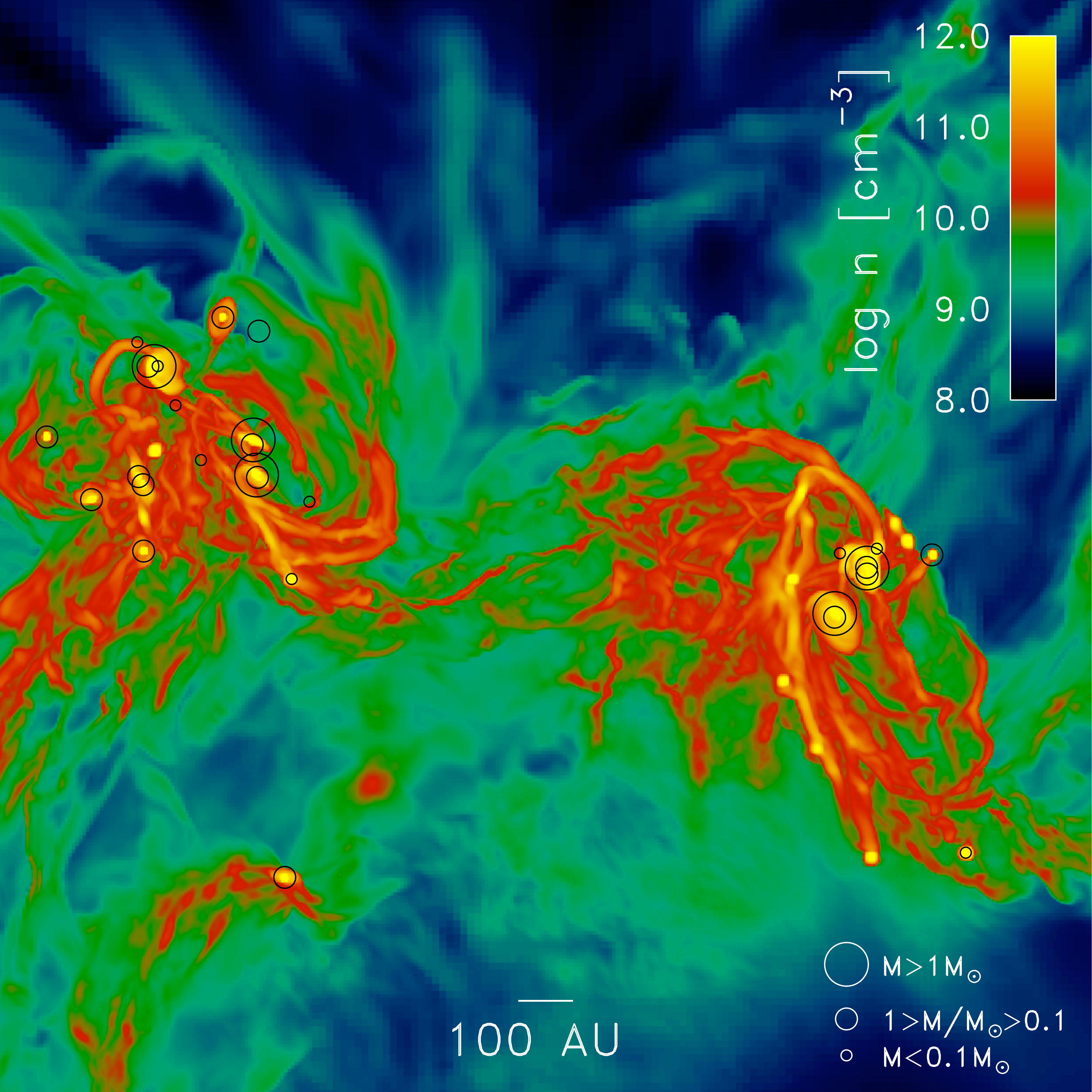}\hspace{0.01em}  &
  \includegraphics[width=1.65in ]{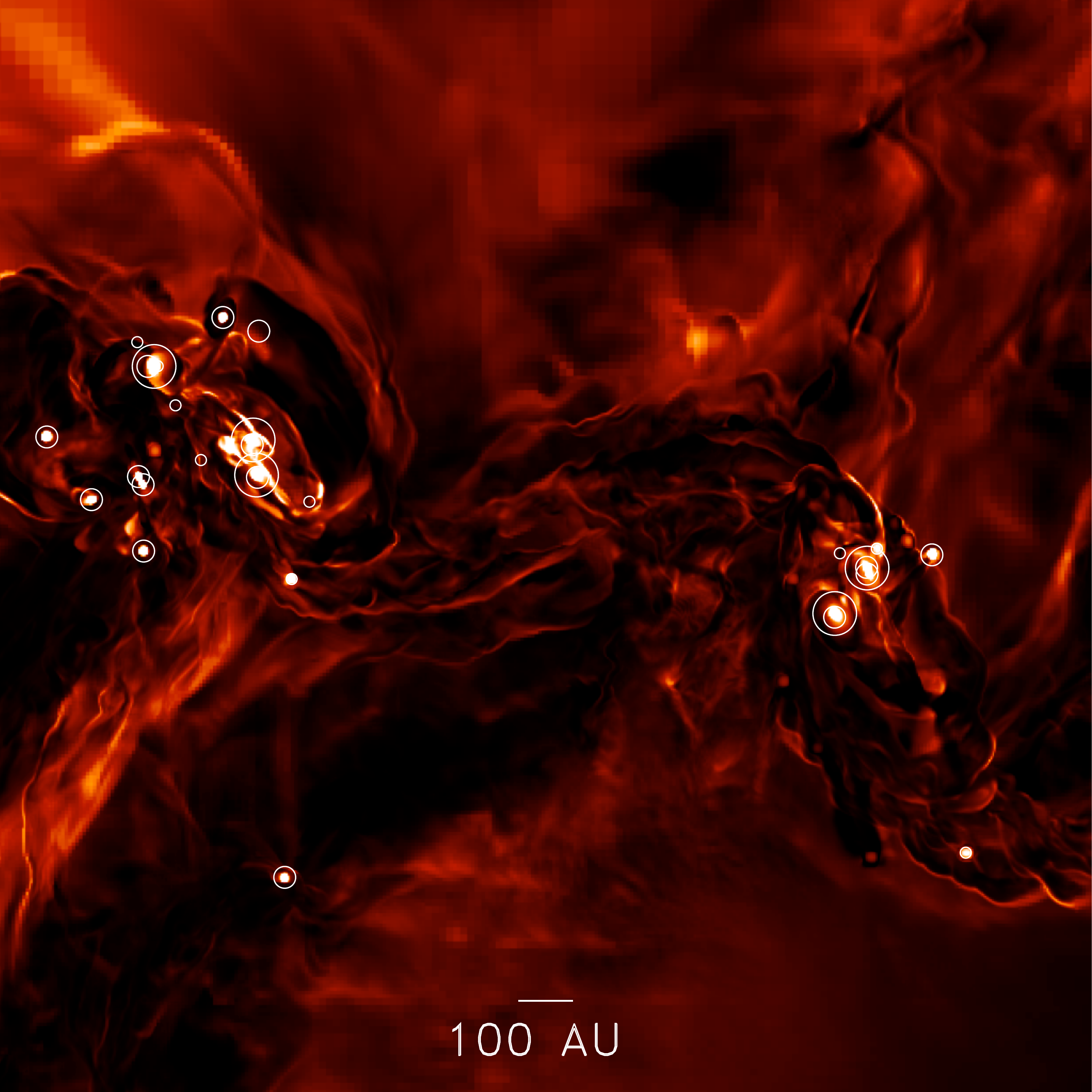} \\
  \includegraphics[ width=1.65in]{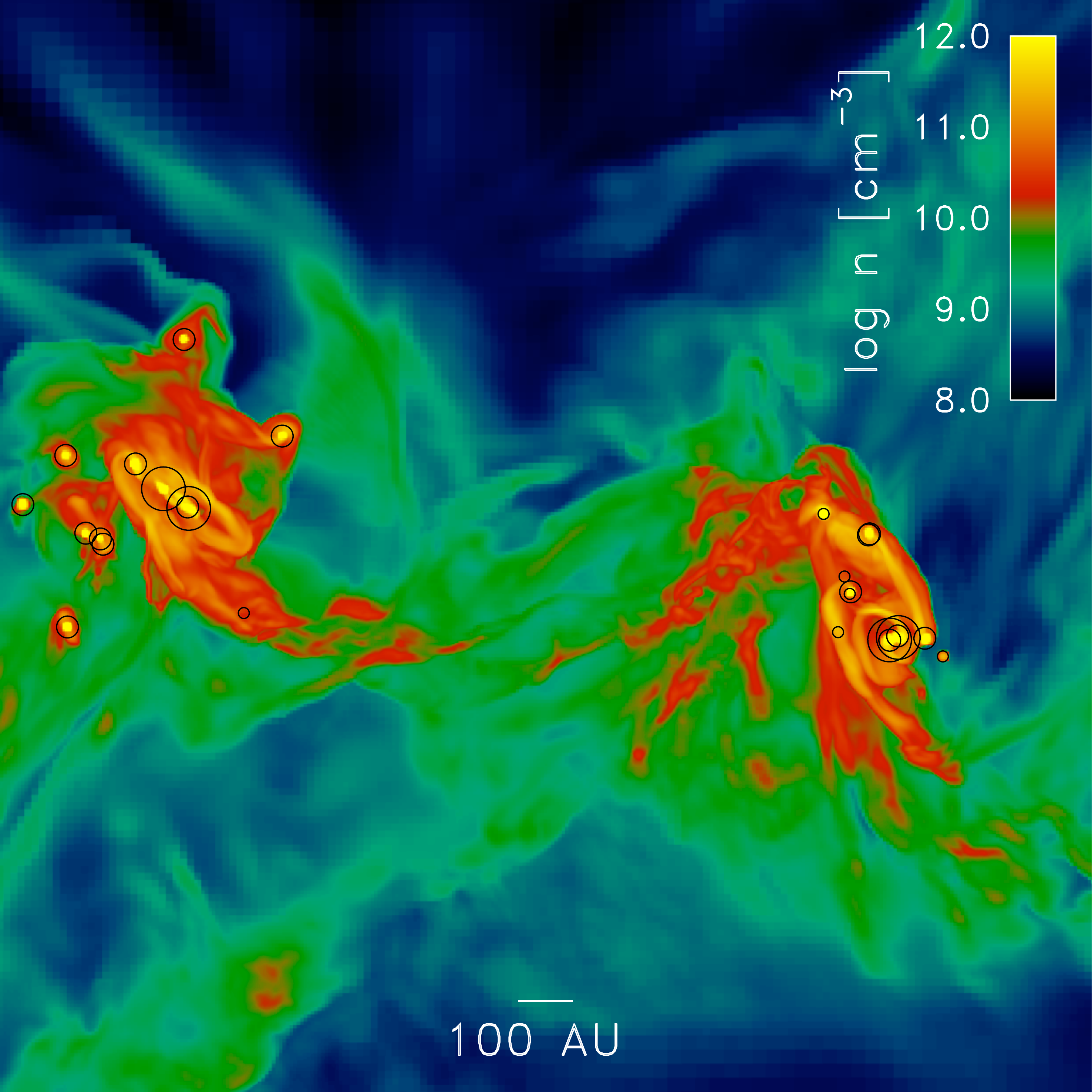}\hspace{0.01em}  &
  \includegraphics[width=1.65in ]{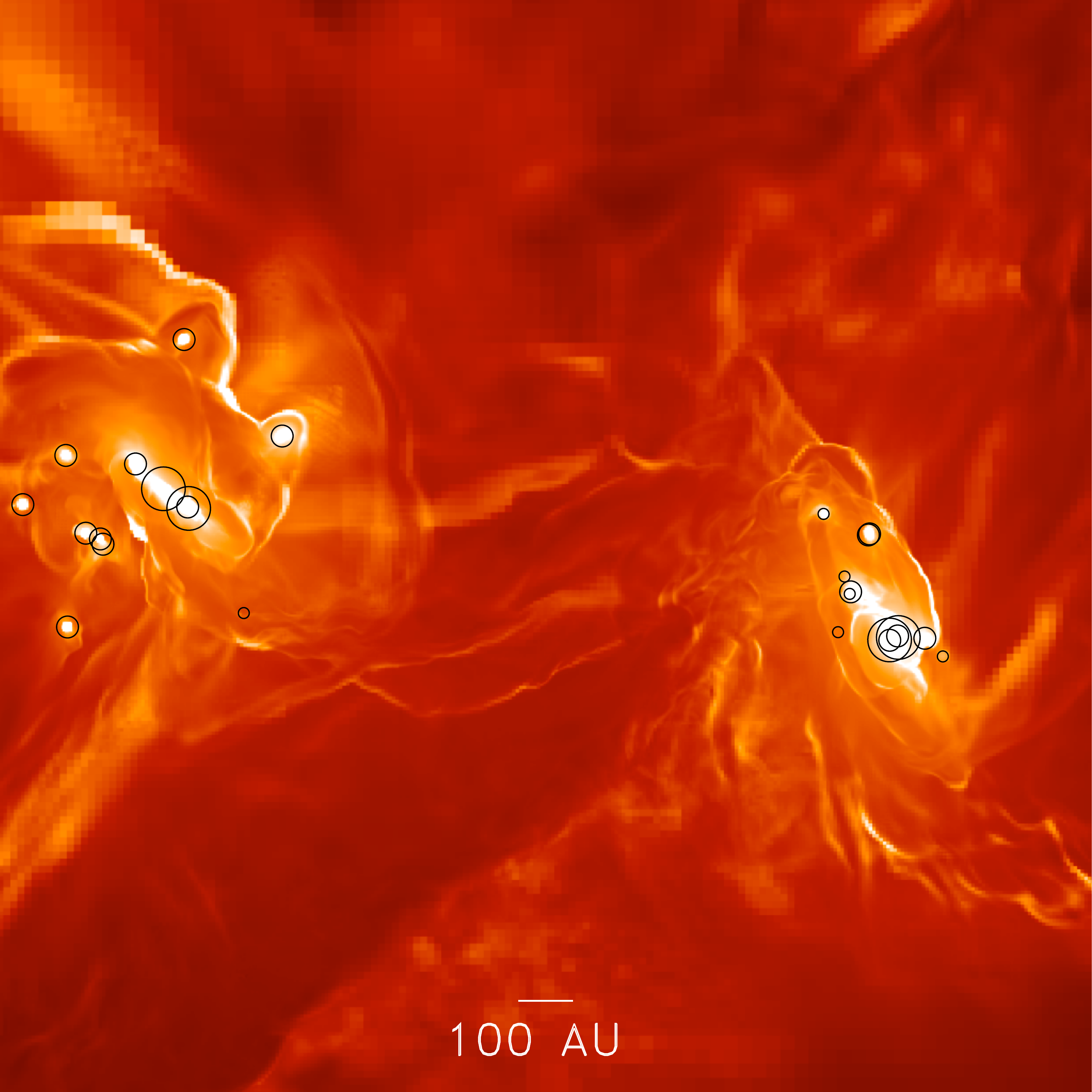} \\
\end{tabular}
%\end{center}
\caption{Mass-weighted line-of-sight projections of gas density (left
  column) and temperature (right column) at the end of \textsc{noheat}
  (top three rows) and at the end of \textsc{heat} (bottom
  row). Circles denote sinks with the size increasing with sink
  mass. In the lower four panels, the smallest circles are sized with the
  sink accretion radius, $r_{\mathrm{acc}}=10\,\au$.}
\label{fig:nofeed_morph}
\end{figure}

We inserted sinks when the density in a cell exceeded
$n_{\mathrm{sink}}=10^{13}\,\cc$, the gas flow was converging $\nabla\cdot\textbf{v}<0$, the
gravitational potential was a local minimum, and a small control
volume around the cell was gravitationally bound.  
In practice, sink formation is preceded by Jeans instability in this highly compressed gas.
Cells within the sink's accretion radius $\racc = 10\,\au =
 2.5\,\Delta x_{\mathrm{min}}$ with hydrogen densities
 $\nh>n_{\mathrm{sink}}$ transferred a fraction
 % were set to transfer a portion
 $(\nh-n_{\mathrm{sink}})/\nh$ of their mass to the sink if the gas
 was gravitationally bound to the sink and had a radial velocity
 directed towards it. Here, $\Delta x_{\rm min}$ is the cell size at
 the highest level of grid refinement.  We did not allow sinks to merge with each 
 other. Sink particle motions were sub-cycled with a leapfrog scheme 
 (for further details, see \citealt{Federrath10} and \citealt{SafranekShrader14}).
 
We utilized the thermodynamical model and non-equilibrium chemical
network described in \citet{SafranekShrader10,SafranekShrader12,SafranekShrader14}, 
now augmented with the dust processes in \citet{Omukai05}. The
dust temperature, $\tdust$, was determined by grain thermal balance in
the presence of thermal emission, heating by the CMB and protostellar 
radiation, and thermal coupling to the gas:
 \begin{align}
 4\sigma_{\rm SB}(\tdust^4 - \tcmb^4)\kappa_{\rm d}(\tdust)&\rho \betaesc = \frac{2\kb (T_{\rm g}-\tdust)n_{\mathrm{d}}}{t_{\mathrm{coll}}} \nonumber \\
 &+ \sum_i\left(\frac{L_{\mathrm{acc},i} }{4\pi r^2_i}\right)\,\kappa_{\rm d}(\tdust)\rho\betaesc ,
 \label{eq:tdust_eqn}
\end{align}
where $\rho$ is the gas density, $n_{\mathrm{d}}$ is the number
density of dust grains, $T_{\rm g}$ and $T_{\rm CMB}$
are the gas and CMB temperatures, and $k_{\rm B}$ and $\sigma_{\rm SB}$ are
the Boltzmann and Stefan-Boltzmann constants. The collision time
between gas and dust particles is $t_{\mathrm{coll}}^{-1} =
\nh\sigma_{\mathrm{d}}\bar{v}_{\mathrm{H}} f$ where
$\sigma_{\mathrm{d}}$ is the average dust grain cross-section,
$\bar{v}_{\mathrm{H}}$ is the average speed of hydrogen nuclei, and
$f\approx0.4$ accounts for non-hydrogenic species
\citep{Schneider06}. The Planck mean opacity of dust grains,
$\kappa_{\rm d}(T)$, was taken from \citet{Semenov03} and we assumed
linear scaling with metallicity. Thermal dust emission, which is a
source of gas cooling through collisional coupling, was attenuated by
a factor $\betaesc = \mathrm{min}(1,\taucont^{-2})$, appropriate for
optically-thick radiative diffusion \citep[e.g.,][]{Masunaga98}. The
continuum optical depth is given by $\taucont = (\kappa_{\rm d} +
\kappa_{\rm g})\rho\lj$ where $\lj$ is the Jeans Length, a local estimate 
for the physical extent of a gravitationally collapsing core, and 
$\kappa_{\rm g}$ is the metal-free gas Planck mean opacity
\citep{Mayer05}. To determine the metal fine-structure line cooling
rates we iteratively calculated consistent line escape probabilities
and level populations \citep[e.g.,][]{Takahashi83,Omukai00}, using a
local estimate of the Sobolev length, $L_{\mathrm{sob}} = \cs /
|\nabla\cdot\textbf{v}|$, to approximate the size of the
shielding region.

The summand in the last term of Equation \ref{eq:tdust_eqn} represents the heating by
the radiation of the $i\mathrm{th}$ protostar located at distance $r_i$ producing accretion
luminosity $L_{\mathrm{acc},i} = GM_{*,i}\dot{M}_{*,i}/R_{*,i}$. By treating sinks 
as the sources of radiation, we took $M_*$ to be the sink
mass and $\dot{M}_*$ to be the sink accretion rate
smoothed over a $10$ yr ($\sim30$ hydrodynamical timesteps)
period. This assumes that the luminosity is dominated by the accretion
luminosity and that all mass accreted by a sink is 
immediately and permanently incorporated into the protostar.  
We calculated the radius of the protostellar photosphere $R_*$ 
with an analytic fit from \citet{Stahler86}, valid independent of 
metallicity for $M_{*}\lesssim3\,\msun$ \citep[e.g.,][]{Hosokawa09}.

 \begin{figure}
\includegraphics[width=0.49\textwidth]{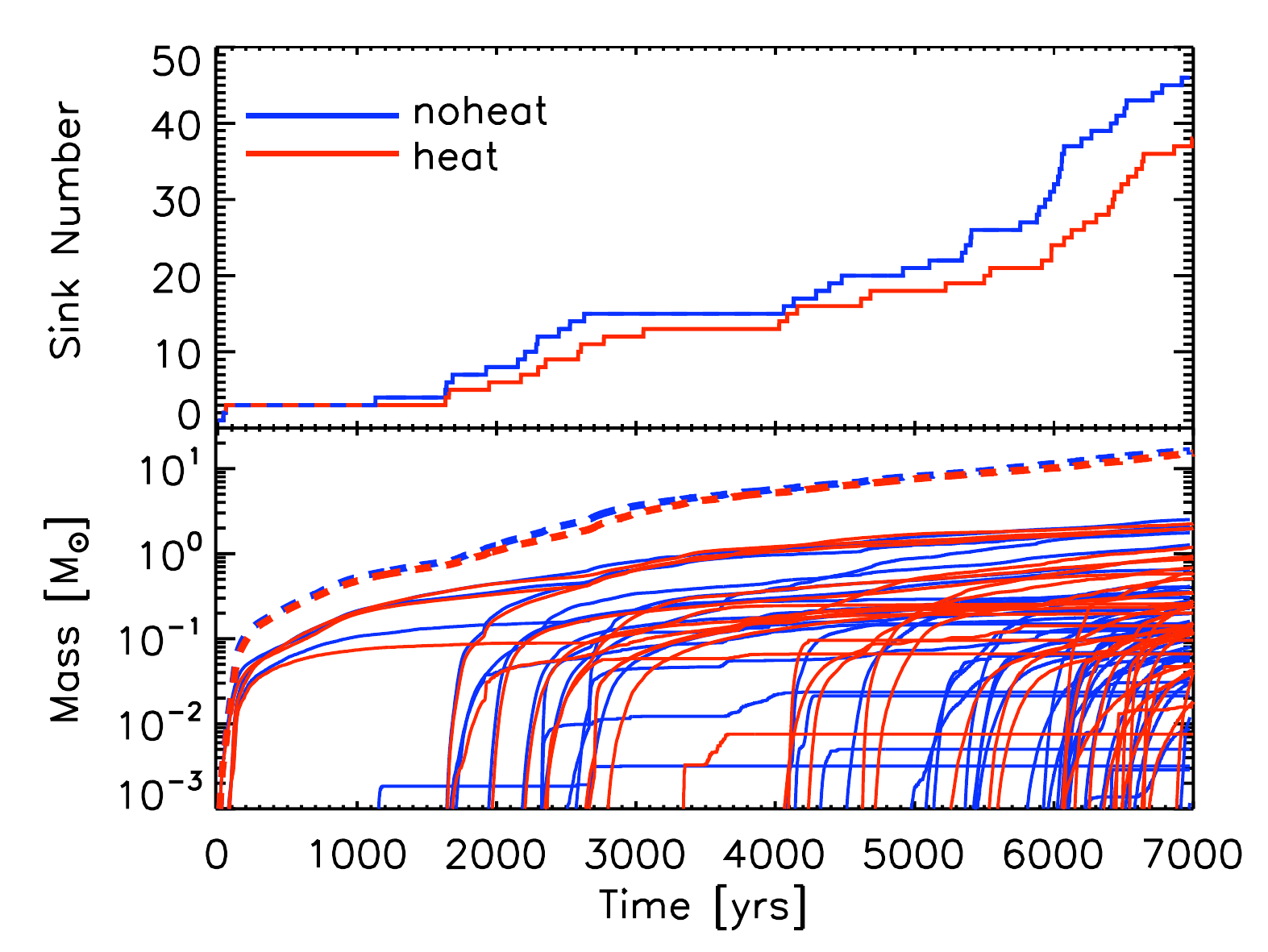}
\caption{Number (upper panel) and masses (lower panel) of sink particles as
  a function of time since first sink formation in
  \textsc{heat} (red lines) and \textsc{noheat} (blue lines).
 In the lower panel, solid
 and dashed
  lines show individual and total sink masses, respectively.}
\label{fig:sink_mass_accrete}
\end{figure}

\section{Results}
\label{sec:results}

We run a simulation \textsc{heat} that includes dust heating by protostellar
radiation (via the last term in Eq.~\ref{eq:tdust_eqn}) and a
reference simulation \textsc{noheat} without heating.
The thermodynamic evolution of gas and dust is shown in Figure \ref{fig:dens_temp}.
As the collapse proceeds, efficient fine-structure line cooling by
[C\,II] and [O\,I] keeps the gas nearly isothermal at $\tcmb\approx40\,\kelvin$ at
densities $\lesssim10^7\,\cc$. Above this density, the lines become
optically thick and the gas heats slightly, but cools back to
$\tcmb$ after reaching densities $\gtrsim 10^9\,\cc$ where gas and
dust collisionally couple. 
%and is minimal at lower densities. 
Isothermal collapse then continues until reaching densities
$\sim5\times10^{11}\,\cc$. At these densities, marking the opacity limit for fragmentation,
 the continuum optical depth due to dust exceeds unity 
 ($\betaesc\lesssim1$, see Eq.~\ref{eq:tdust_eqn}), 
dust-cooling loses its efficacy, and the gas
begins to evolve adiabatically.  
%This transition, marking the opacity limit
%for fragmentation, is the result of the energy input from gravitational collapse exceeding
%the maximum rate at which energy can be radiated \citep[e.g.,][]{Masunaga99}. 
The effect of protostellar dust heating is minimal, mainly 
resulting in higher gas temperatures at densities $\gtrsim 10^8\,\cc$ 
and a slight suppression of sink formation, 
consistent with the findings of \citet{Omukai10}.

 %when the energy input from gravitational collapse exceeds
%the maximum rate at which energy can be radiated \citep[e.g.,][]{Masunaga99}.

When the gas reaches densities $\sim 10^{13}\,\cc$ sink formation is
possible based on the conditions described in Section
\ref{sec:method}. The first sink forms $4.3\times10^4$ yr
after the beginning of the simulations
and both are run for $7000$ yr after this point.
Figure \ref{fig:nofeed_morph}
shows mass-weighted density and temperature projections at the end of both simulations.
Sinks are forming over an extended $\sim10^4\,\au$-long
filamentary structure, with thickness $\sim1000\,\au$ and
density $\sim10^8\,\cc$.  The structure is produced by a large-scale, supersonic
colliding flow and is undergoing global gravitational collapse.
We identify two sites of sink formation: in locally fragmenting
filamentary structures, and in rotating, quasi-virialized disky
flows produced by progressive global gravitational collapse of the
filaments (as seen, respectively, in row $2$ and rows $3$ and $4$ of Fig.~\ref{fig:nofeed_morph}).

 \begin{figure}
%\includegraphics[scale=0.5, clip, trim = 20 5 5 5 ]{fig2}
%\begin{center}
\centerline{\includegraphics[width=0.42\textwidth]{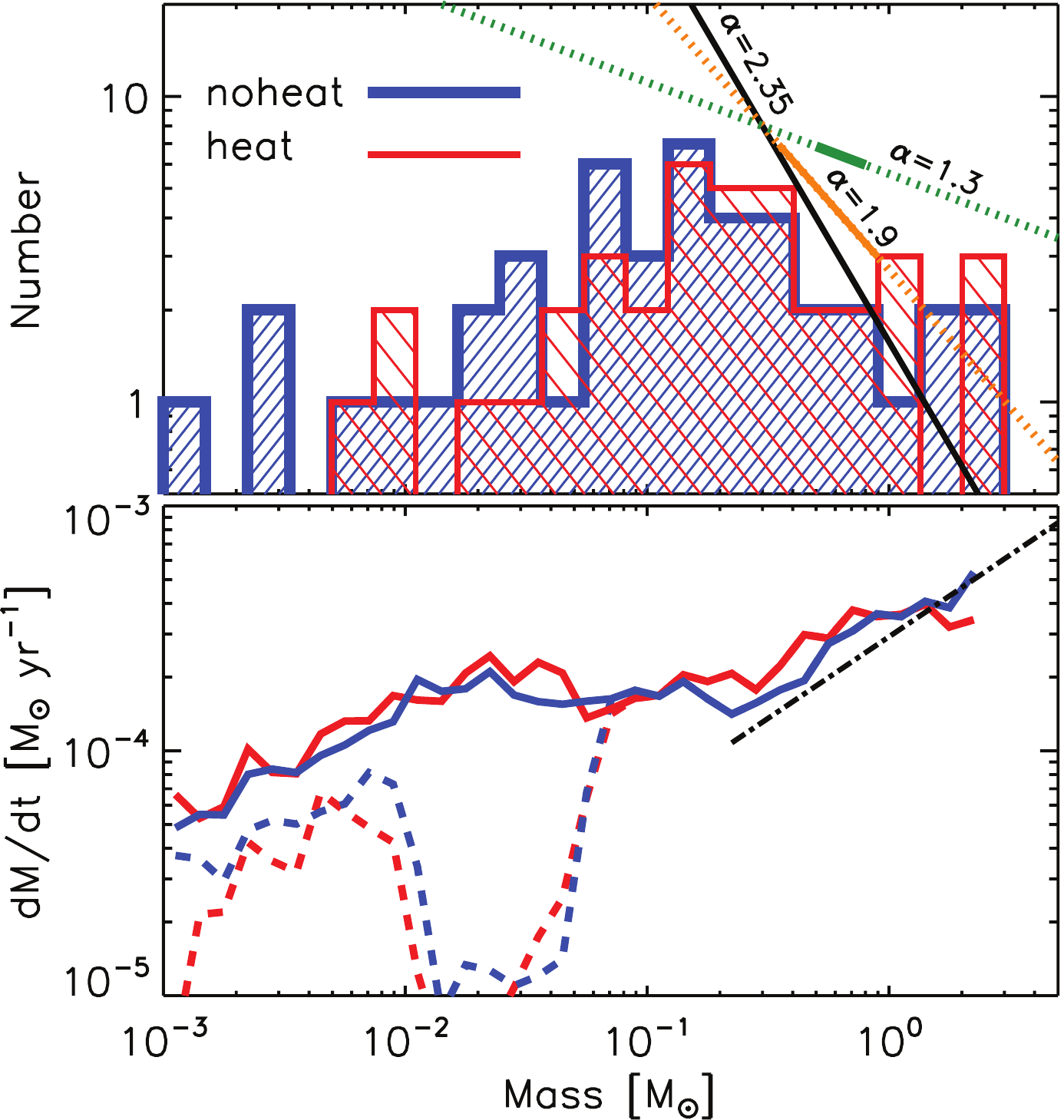}}
%\end{center}
\caption{Upper panel: Sink particle mass function in \textsc{heat} (red line) and
  \textsc{noheat} (blue line). Dashed lines show power-law IMF slopes 
  $\alpha=2.35$ (Salpeter), $\alpha=1.9$ (in the Small Magellanic
  Cloud; \citealt{Kalirai13}), and $\alpha=1.3$ (the UFD Leo~IV;
  \citealt{Geha13}). The mass range over which the power-law slope was
  measured is indicated with a solid line.  Lower panel:
average sink accretion rate as a function of sink mass, including all
sinks (dashed lines) and only accreting sinks (solid lines).  The
dot-dashed line shows the scaling ${\dot M}_\star\propto M_*^{2/3}$.}
\label{fig:sink_mass_spectrum}
\end{figure}

Figure \ref{fig:sink_mass_accrete} shows the total number of sinks and their
individual masses.  In \textsc{noheat}, $46$ sinks formed with a total mass of 
$16\,\msun$. In \textsc{heat}, $37$ sinks formed with a total mass of $15\,\msun$.  
The average sink mass in both simulations is 
$M_{*,\mathrm{tot}}/N_{\mathrm{tot}}\approx0.3-0.4\,\msun$ and median
mass is $\approx0.13\,\msun$.  Sink binaries form in both
simulations and we observe that binary
interactions eject a number of low-mass sinks ($<0.08\,\msun$) from the star forming
region with velocities $3-7\,\textrm{km}\,\textrm{s}^{-1}$. Figure
\ref{fig:sink_mass_spectrum} shows the average sink accretion rate as a function of 
mass.  Sinks that are not ejected accrete at an average
individual accretion rate of $2\times10^{-4}\,\msunperyr \approx \mathrm{few}\times\cs^3/G$ while their
mass is in the range $0.01\,\msun< M_*< 0.3\,\msun$.  At higher masses,
the accretion rate increases approximately as $\dot{M}_*\propto
M_*^{2/3}$ as sinks accrete from within the extended gaseous disks in which they are 
embedded.  This trend continues up to the highest sink 
mass in \textsc{noheat}, but in \textsc{heat} the increase of accretion rate
with mass levels above $M_*\sim 0.7\,\msun$.

Figure \ref{fig:sink_mass_spectrum} also shows the sink mass
function at the end of both simulations. Because sinks do not
instantaneously accrete their initial gravitationally unstable Jeans mass
$\gtrsim 0.01\,\msun$, the simulations contain some sinks with very low 
masses that cannot be interpreted as fully fledged proto- or sub-stellar objects.
The mass spectrum $dN/d\ln M_*$ in both simulations
exhibits a broad peak at $M_*\sim (0.05-0.4)\,\msun$.  The most
massive sinks $M_*\approx 2.5\,\msun$ are located close to the centers of the densest disky
structures (Fig.~\ref{fig:nofeed_morph}), already indicating hints of primordial 
segregation in the proto-stellar cluster.

\section{Discussion and Conclusions}
\label{sec:discussion}

Studies that investigate the control of gravitational fragmentation in
collapsing gas clouds using one-zone models,
 \citep[e.g.,][]{Omukai05,Omukai10,Schneider06} assume that the 
fragmentation mass scale is set by the interplay of monolithic 
gravitational collapse and thermodynamics, neglecting large-scale, multi-dimensional 
gas motions. They predict that at the metallicity we
consider, the characteristic fragmentation mass
should be $\sim0.5\,\msun$ as a result of gas-dust coupling at high
densities $\gtrsim10^8\,\cc$. Since the Jeans mass $\mj\propto\rho^{-1/2}$ and 
the compression at Mach number ${\cal M}$ results in a factor of ${\cal M}^2$ density
enhancement, the characteristic mass should be revised downward by
a factor $\sim1/{\cal M}$. With the Mach number of the inflow into the fragmenting structure 
in the simulations being ${\cal M}\sim3$, this revises the 
prediction of the one-zone models for the characteristic fragmentation 
mass to $\sim0.1\,\msun$, consistent with the peak of our sink mass function.
% The peak of our measured sink mass function is roughly consistent 
%with this prediction, suggesting that gas-dust coupling does
%likely play a role in setting the characteristic fragmentation, and thus stellar, mass scale.
%Studies that investigate the
%control of gravitational fragmentation in collapsing gas
%clouds assuming large-scale, multidimensional
%motions can be neglected \citep[e.g.,][]{Omukai05,Omukai10,Schneider06}.  
%In these models, the fragmentation mass scale is set by the
%interplay of monolithic gravitational collapse and thermodynamics. 
%They predict that at the metallicity we
%consider, the characteristic fragmentation mass
%should be $\sim0.5\,\msun$ as a result of gas-dust coupling at high
%densities $\gtrsim10^8\,\cc$ \citep[e.g.,][]{Schneider10}. Since this is 
%similar the peak of our measured sink mass function, it seems that gas-dust 
%coupling does indeed play a role in setting the characteristic fragmentation, 
%and thus stellar, mass scale.
%The presence of large-scale compressive motions in
%our simulations implies that these models may not be directly
%applicable. 
Models for gravitational fragmentation that do 
account for the underlying supersonically turbulent gas flow
normally assume fully developed,
statistically-homogeneous turbulence,
as in molecular clouds, but potentially inapplicable in
the present context, and idealized
thermodynamic evolution \citep[e.g.,][]{Padoan02}.\\
\indent A more definitive theoretical point of reference are 
simulations that resolve fragmentation into individual stars, albeit
in a variety of contexts widely different from ours (e.g., Milky Way disk, metal-free or
ultra-low-metallicity minihalos, etc.).
\citet{Clark08} and \citet{Dopcke13} studied the
onset of sink formation in clouds initialized with imposed turbulent velocity
fluctuations at very low metallicities ($\leq 10^{-4}\,\zsun$),
and followed the evolution of the cluster for only up to few hundred
years.  They identified gas-dust thermal coupling as the trigger of
the conclusive fragmentation episode in this regime.
\citet{Bate12} and \citet{Krumholz12} carried out longer-duration integrations, also
in clouds with imposed turbulent density fluctuations, designed to model the
formation of small star clusters at $Z=Z_\odot$, both
including protostellar radiative feedback. They found that the peak and 
shape of the resulting IMF are
delicate functions of protostellar feedback and of how
turbulence is introduced \citep{Krumholz12}.\\
\indent An isolated sink will accrete its initial Jeans
mass, but a sink interacting with other matter (e.g., being subject to
a tidal field, becoming a member of a binary system, or being ejected via a
$\geq3$ body interaction) may fail to attain the Jeans mass. The
most massive sinks at the end of the simulations owe their status to
sustained accretion, at a typical average rate of
$\sim4\times10^{-4}\,\msunperyr$.  The sites of this sustained accretion are the disky
structures, mentioned in Section \ref{sec:results} and previously reported by \cite{Clark08}, 
in which shearing seems as important as turbulence. Various theoretical models
for continued accretion after a core has accreted its initial
Jeans mass typically predict a dependence of the mass accretion rate on
the protostellar mass $\dot{M}_* \propto M_*^\xi$ \citep[e.g.,][]{McKee03}.  
Tentatively we find $\xi\approx 2/3$ at least in the range of stellar masses in which
our two simulations agree. At higher
masses $>1\,\msun$ local protostellar heating in an increasingly crowded
environment may be depressing 
Bondi-Hoyle accretion onto the most massive sinks. 
%As long as they avoid ejection, massive
%sinks will continue accreting, but
%protostellar heating in an increasingly crowded 
%protostellar
%environment will become a more important factor influencing
%the character of
%long-term accretion.  
If the long-term accretion rate does remain at
$\sim2\times10^{-4}\,\msun\,\yr^{-1}$, the cluster will form a $\sim20\,\msun$ star in
$\sim10^5\,\yr$ with the potential to produce an H\,II region that
could truncate star formation.\\
\indent The present simulation resolves
small-scale fragmentation in only one of several (9
forming within $\sim3\,\textrm{Myr}$)
pre-stellar clumps satisfying the conditions for gravitational collapse and
star formation in the parent cosmological simulation. 
The other clumps would likely
undergo a similar small scale evolution, implying that the
total number of sinks, and thus stars, in a single episode of star
formation could be an order of magnitude larger.  Moreover, a single
clump, like the one followed here for $7000\,\yr$ after initial
sink formation, will continue infall on a dynamical time $\sim
3\times10^4\,\yr$, roughly five times longer than integrated here,
producing $\sim 5\times 40=200$ stars in one pre-stellar clump, or $\sim2000$
stars in the whole halo. This value is interestingly consistent with the stellar
masses of the faintest UFDs, Bo\"otes II, Segue I \& II, and Willman I. The surrounding 
halo at the time of
star formation contains $\approx2\times10^5\,\msun$ of dark matter within
a reference radius of $30\,\pc$, furthermore consistent with the stellar-kinematics-inferred 
dynamical masses of these UFDs \citep{McConnachie12}.\\
\indent The coarse-grained sinks representing pre-stellar
clumps in \citet{SafranekShrader14} and the fine-grained ones
representing individual stars in the present work occupy a
highly compact
spatial region ($\sim1\,\pc$) in which the gravitational potential is strongly
baryon-dominated, whereas the stars in UFDs occupy much more extended,
dark-matter-dominated regions.  Long term evolution 
will inevitably convert the compact
baryon-dominated star cluster into a more diffuse, highly dark-matter-dominated 
one.  We already observe ejections of low-mass 
sinks by dynamical interactions, similar to \citet{Bate12}. Given
the low total number of stars in the cluster
ejections will undoubtedly continue, extending to stellar mass objects,
as long as the cluster remains
baryon-dominated.  If the ejected stars do not escape the host dark
matter halo they will predominantly reside at large, dark-matter-dominated
galactocentric radii.  Stellar mass loss will further
increase the mass-to-light ratio.\\
\indent The stellar IMF in two UFDs measured in the mass range
$0.5\,M_\odot<M_*<0.8\,M_\odot$ is shallower, $dN/dM_*\propto
M_*^{-\alpha}$ with $\alpha = 1.3 - 1.4$
\citep{Geha13}, than the IMF of the Milky Way  ($\alpha=2.3$;
\citealt{Kroupa01}) and the Small Magellanic Cloud ($\alpha=1.9$; \citealt{Kalirai13}) 
in a similar mass range.  While we emphasize that the
stellar IMF in our simulations is still evolving at the end of each simulation, we do find that for
$M_* >0.1\,\msun$, a truncated Pareto distribution maximum likelihood
estimate \citep{Aban06} yields $\alpha=1.26\pm0.05$ (\textsc{heat})
and $\alpha=1.41\pm0.09$ (\textsc{noheat}), consistent with the
\citet{Geha13} values.  \\
\indent It should be kept in mind that at present it is not clear whether UFDs
started forming low-mass stars before reionization, where they were subject
only to a non-ionizing $\htwo$-dissociating UV background, or if their
first low-mass stars formed only in a patch of the Universe reionized
by more massive neighboring halos. In the latter case, we expect a
similar outcome in a more massive host
halo, though possibly modified due to higher inflow velocities and larger 
clump accretion rates. Furthermore,
an object already forming stars at $z\approx14$ could have ended
up being incorporated into the central part of the Milky Way, rather
than remaining an orbiting satellite. One
central issue to be resolved is which process terminated star
formation in UFDs: internal feedback from star formation,
photoevaporation by reionization, or ram pressure stripping during
infall into a more massive Milky Way progenitor halo.\\
\indent We will report on a more comprehensive set of simulations and test cases, including
a detailed analysis of the role of turbulence, in forthcoming publications.
We are entering a period of rapid discovery, where ever more realistic simulations and
comprehensive stellar archaeological surveys promise a deeper understanding of
our ancient cosmic origins.

\section*{Acknowledgments}

The authors thank Christoph Federrath, Marla Geha, John Wise, 
Kazu Omukai, Raffaella Schneider, Anna Frebel, and many others for useful discussions. 
We also thank an anonymous referee for helpful comments.
The FLASH code was in part developed by the DOE-supported Flash Center
for Computational Science at the University of Chicago. The authors acknowledge the Texas Advanced Computing Center at The University of Texas at Austin for providing HPC resources under XSEDE allocation TG-AST120024. CSS is grateful for support provided by the NASA Earth and Space Science Fellowship (NESSF) program. This study was supported in part by NSF grant AST-1009928 and by the NASA grant NNX09AJ33G.

\footnotesize{
\bibliographystyle{mn2e_fixed}
\bibliography{dust_collapse}

\IfFileExists{\jobname.bbl}{}
 {\typeout{}
  \typeout{******************************************}
  \typeout{** Please run "bibtex \jobname" to optain}
  \typeout{** the bibliography and then re-run LaTeX}
  \typeout{** twice to fix the references!}
  \typeout{******************************************}
  \typeout{}
 }
}

\label{lastpage}

\end{document}